\documentclass[11pt]{article}

\usepackage[margin=1in]{geometry}
\usepackage{booktabs}
\usepackage{tabularx}
\usepackage{array}
\usepackage{graphicx}
\usepackage{hyperref}
\usepackage[numbers]{natbib}

\title{Contractual Skills: A GovernSpec Design Framework for Enterprise AI Agents}

\author{
Ting Liu\\
SymbolicLight Research\\
Foshan, Guangdong, China\\
\texttt{research@symboliclight.com}
}

\date{May 2026}

\begin{document}

\maketitle

\begin{abstract}
Skills have become a practical packaging mechanism for agent instructions, workflows, scripts, and reference materials. In enterprise settings, however, a skill is often expected to do more than trigger relevant context. It must also express input boundaries, permissions, human approval points, evidence requirements, output contracts, quality criteria, verification steps, and handoff rules. When these controls are written as informal prose, they are difficult to review, reuse, test, or connect to runtime guardrails. This paper proposes \emph{contractual skills}, a design framework that organizes \texttt{SKILL.md} files with GovernSpec-style task contract fields while preserving the lightweight discovery and progressive-loading properties of skills. The framework positions a skill as a readable task contract rather than as a free-form prompt fragment. It also clarifies the boundary between contractual skills, GovernSpec YAML contracts, Model Context Protocol (MCP) surfaces, tool adapters, runtime guardrails, tracing, and evaluation systems.

We evaluate the framework with three offline empirical studies. The first text-generation experiment covers three enterprise skills, fifteen synthetic tasks, four instruction conditions, and eight generation models, producing 960 outputs and 1680 cross-judge score records. Contractual skills receive higher mean model-judge scores than the no-skill and minimal-skill baselines on all eight models. Compared with a plain expanded skill that contains similar information but lacks contract fields, contractual skills are slightly higher on six models and slightly lower on two models, with small differences. The second study is a market-validated skill A/B expansion: eight public skills are compared with contractual rewrites across forty-eight synthetic tasks, six generation models, two repeats, 1152 outputs, and two complete judge files. In this setting, contractual skills raise mean quality from 4.692 to 4.914 and reduce critical-error rate from 0.083 to 0.013. The third study is an offline tool-calling challenge with eight models and 192 simulated tool-call records. Skills usually reduce high-risk tool attempts, but the effect varies by model; contractual skills do not replace tool-level guardrails. The results suggest that contractual skills are most useful as a governance layer for making task intent, boundaries, and acceptance criteria explicit, rather than as a standalone safety mechanism.
\end{abstract}

\paragraph{Code and data availability.}
The public replication package, including synthetic tasks, skill variants, prompts, model outputs, tool-calling transcripts, scoring records, analysis scripts, and reusable templates, is available at \href{https://github.com/SymbolicLight-AGI/contractual-skill}{\texttt{SymbolicLight-AGI/}\allowbreak\texttt{contractual-skill}}.

\section{Introduction}

Large language model applications are moving from single-turn question answering toward agent systems that read files, call tools, retrieve knowledge, generate artifacts, and coordinate multi-step workflows. In enterprise settings, agents increasingly participate in sales follow-up, pre-sales solution drafting, contract review, delivery management, customer success, knowledge operations, finance checks, and code review. These workflows combine business context, organizational rules, customer-facing commitments, privacy constraints, external systems, and human approval. They therefore require more than fluent text generation.

Skills are a common way to package reusable agent capabilities. A skill usually consists of a directory with a \texttt{SKILL.md} file. Its frontmatter provides discovery metadata such as \texttt{name} and \texttt{description}; the body provides task guidance, workflows, references, and sometimes scripts. Anthropic's Agent Skills documentation describes skills as modular packages of instructions, executable code, and resources that the model can use when relevant \citep{anthropicSkills}. This design is attractive because it is local, inspectable, and compatible with progressive loading.

Yet enterprise skills often need to express more than task advice. A sales-growth skill should not merely tell an agent how to analyze a customer account; it should also prevent the agent from promising prices, delivery dates, final discount terms, or guaranteed outcomes. A finance-contract skill should separate facts from assumptions, identify approval points, and make clear when a draft is not a binding commitment. A code-review skill should prioritize behavioral risks, avoid unrelated edits, and state test gaps. In other words, enterprise skills must encode task goals, behavioral boundaries, and acceptance criteria.

This paper proposes \emph{contractual skills}. A contractual skill is still a \texttt{SKILL.md} file, but its body is organized as a task contract. The recommended sections include goal, audience, inputs, context, workflow, permissions, human gates, constraints, evidence, output, quality bar, verification, and handoff. The term is motivated by GovernSpec, a local-first contract compiler for AI task governance. In prior work, \citet{liu2026governspec} introduced GovernSpec as a runtime-independent workflow that validates a single task-governance contract, compiles it into multiple agent artifacts, imports existing artifacts, and checks outputs with offline assertions. This paper focuses on the skill layer: how to bring the same contract-oriented discipline into the Markdown files that agents actually load as capabilities.

The central claim is modest. Contractual skills do not make a model inherently safe, nor do they replace runtime permission checks. Their value is to make the task contract explicit and reviewable. They help the model, the maintainer, and the evaluator refer to the same fields when deciding what inputs are required, which actions are allowed, what evidence is sufficient, what output shape is expected, and when work must be handed off or paused for human review.

This paper makes five contributions:

\begin{enumerate}
  \item It defines contractual skills as a GovernSpec-style organization pattern for \texttt{SKILL.md}.
  \item It proposes a field model for enterprise skills, including inputs, permissions, human gates, evidence, output, verification, and handoff.
  \item It clarifies the boundary between contractual skills, GovernSpec YAML, MCP, tool adapters, runtime guardrails, tracing, and evaluation.
  \item It reports a multi-model text-generation experiment with 960 outputs and 1680 cross-judge records.
  \item It reports a market-validated skill A/B expansion with eight public skills, 1152 outputs, and two complete judge files.
  \item It reports an offline simulated tool-calling challenge with 192 records and discusses where skill-level contracts help or fail.
\end{enumerate}

\section{Background and Related Work}

\subsection{Agent skills and progressive loading}

Agent skills package instructions, scripts, and resources into reusable units. The frontmatter helps the agent decide when to load the skill, while the body and auxiliary files provide task-specific guidance. This is useful for document workflows, domain-specific analysis, coding conventions, and enterprise procedures. The core benefit is context economy: the agent does not need to carry every instruction in the main prompt.

However, a lightweight skill format can also hide important governance information. Rules about customer commitments, privacy, human approval, source quality, and output acceptance may be scattered across prose paragraphs. When a skill library grows across sales, finance, delivery, risk, customer success, and engineering, maintainers need a way to audit whether each skill contains the relevant boundaries.

Recent empirical and security studies make this governance problem more concrete. \citet{ling2026agentskills} analyze 40,285 publicly listed Claude skills and report ecosystem redundancy, concentration in software engineering workflows, and non-trivial safety risks including state-changing or system-level actions. \citet{li2026secureagentskills} study the Agent Skills lifecycle and identify structural risks such as weak data-instruction separation and persistent trust assumptions. \citet{duan2026skillattack} show that even apparently benign skills can expose exploitable vulnerabilities under adversarial prompting. These studies motivate the need for skill-level structure, reviewability, and runtime separation, which are central concerns of contractual skills.

\subsection{GovernSpec and artifact-level governance}

GovernSpec describes AI tasks as explicit YAML contracts. A contract can specify a task goal, permissions, constraints, confirmation gates, output requirements, and deterministic acceptance tests. Liu's SSRN paper frames GovernSpec as runtime-independent artifact-level governance: teams can maintain one governance contract and compile it into downstream artifacts without modifying each agent runtime \citep{liu2026governspec}.

Contractual skills reuse this idea but target a different artifact. GovernSpec YAML is a structured source of truth that can be validated and compiled. A contractual skill is a human-readable instruction artifact that an agent can load directly. The two can be used independently, but the strongest workflow is to let high-risk skills derive from or remain aligned with a GovernSpec contract.

\subsection{Tool calling, guardrails, and tracing}

Tool calling allows a model to interact with external systems through structured calls. OpenAI describes function calling as a way for models to interface with external systems and access data outside the model itself \citep{openaiFunctionCalling}. In agent systems, tool calls may be combined with handoffs, state, guardrails, and traces. Guardrails can check or block inputs, outputs, or tool calls \citep{openaiGuardrails}. Tracing records model generations, tool calls, handoffs, guardrails, and custom events \citep{openaiTracing}.

Contractual skills do not replace these runtime mechanisms. A skill can state that a discount approval must not be executed without human confirmation, but the tool adapter must still enforce that policy. The skill tells the model and maintainer what should happen; the adapter and guardrail determine what can happen.

\subsection{MCP and structured capability exposure}

The Model Context Protocol (MCP) provides a standardized way for servers to expose prompts, resources, and tools to clients \citep{mcpPrompts}. MCP is a runtime-facing protocol surface. A contractual skill is an instruction and governance surface. They are complementary: a contractual skill can describe when to use MCP tools, what inputs are required, and when a call should be blocked or escalated.

\subsection{Reasoning, acting, and evaluation}

ReAct showed that language models can interleave reasoning traces and task-specific actions, allowing the model to update plans based on observations \citep{yao2023react}. Enterprise agents need a similar loop, but with organizational constraints. If evidence is insufficient, the agent should say so. If an action is high-risk, it should pause. If the task crosses a role boundary, it should hand off.

Evaluation also needs to be multidimensional. HELM argues for broad, transparent evaluation across scenarios and metrics \citep{liang2023helm}. For enterprise agents, a single accuracy score is insufficient. We therefore evaluate structure, risk control, evidence quality, output usability, handoff clarity, maintenance consistency, critical errors, and tool-call behavior.

\section{Problem Definition}

\subsection{Limits of informal skills}

An informal skill usually contains discovery metadata and prose guidance. This works for low-risk tasks, but enterprise settings expose five recurring limitations.

First, input boundaries are unclear. A skill may say ``read the customer context'' without defining required inputs, optional inputs, missing-data behavior, or privacy class. The agent may produce confident recommendations from incomplete context.

Second, permissions are not explicit. A skill may mention that the agent should not promise a price or delete files, but such rules are buried in prose. When tools include email, CRM updates, contract editing, file writes, or deployment actions, scattered reminders are not enough.

Third, human approval points are unstable. Price commitments, discount approval, production changes, external messages, and destructive operations often require human confirmation. Without an explicit human-gates section, the model may fail to detect the pause point.

Fourth, evidence policy is weak. Research, risk review, and customer analysis require a distinction between facts, assumptions, inferences, and recommendations. Without a dedicated evidence section, unsupported certainty becomes more likely.

Fifth, output acceptance depends on tacit human expectations. A skill may ask for a customer profile, risks, and next steps, but not define required sections, schema, word limit, or verification criteria. Outputs then drift across tasks and models.

\subsection{Enterprise governance boundaries}

Enterprise agents operate inside organizational workflows. Their outputs can influence customer expectations, contract scope, finance decisions, delivery planning, and engineering responsibility. A skill therefore needs to define at least five boundaries:

\begin{itemize}
  \item customer commitment boundaries, such as price, scope, delivery time, and guaranteed outcomes;
  \item privacy and data boundaries, such as contact information, financial data, internal project issues, and sensitive credentials;
  \item evidence boundaries, such as facts, assumptions, inference, and unsupported claims;
  \item tool boundaries, such as read-only tools, draft-generating tools, external write tools, and destructive tools;
  \item role boundaries, such as when sales should hand off to pre-sales, finance, risk, delivery, or engineering.
\end{itemize}

\subsection{From prompt fragments to organizational assets}

When a team has only a few skills, informal Markdown may be enough. When a company maintains skills across many business lines, skills become organizational assets. Assets must be inspectable, reviewable, reusable, and testable. Contractual skills put governance information in stable sections so that maintainers can quickly check whether a skill has input requirements, permission boundaries, evidence policy, output expectations, and verification criteria.

\section{The Contractual Skill Framework}

\subsection{Design principles}

Contractual skills follow six principles.

\begin{enumerate}
  \item Preserve the lightweight discovery mechanism of skills. The frontmatter remains small and trigger-oriented.
  \item Use fields to clarify execution boundaries, not to create bureaucracy. Not every skill needs every field.
  \item Make high-risk rules explicit. Permissions, human gates, evidence, and handoff should not be hidden in long prose.
  \item Separate skill instructions from runtime enforcement. Skills express intent; adapters, guardrails, and permissions enforce behavior.
  \item Support gradual adoption. Teams can start by rewriting existing skills, then add GovernSpec YAML and offline tests for high-risk workflows.
  \item Support evaluation. Fields should map to checks such as required sections, forbidden commitments, privacy leakage, uncertainty marking, and handoff completeness.
\end{enumerate}

\subsection{Field model}

Table~\ref{tab:fields} summarizes the proposed field model.

\begin{table}[h]
\centering
\small
\begin{tabularx}{\linewidth}{p{0.23\linewidth} X X}
\toprule
Field & Purpose & Governance role \\
\midrule
\texttt{When To Use} & Applicability and non-applicability & Reduces mis-triggering \\
\texttt{Goal} & Task objective and completion state & Reduces goal drift \\
\texttt{Audience} & Target reader or user & Controls tone and granularity \\
\texttt{Inputs} & Required and optional inputs, paths, privacy class & Prevents confident output from missing data \\
\texttt{Context} & Facts, assumptions, glossary & Reduces context confusion \\
\texttt{Workflow} & Execution steps & Improves process consistency \\
\texttt{Permissions} & Allowed and forbidden tools, files, network, and external actions & Reduces overreach \\
\texttt{Human Gates} & Conditions requiring human confirmation & Controls high-risk actions \\
\texttt{Constraints} & Hard rules and prohibitions & Reduces policy violations \\
\texttt{Evidence} & Source requirements and uncertainty marking & Reduces unsupported certainty \\
\texttt{Output} & Format, language, sections, schema, length & Stabilizes deliverables \\
\texttt{Quality Bar} & Criteria for a good result & Improves usefulness \\
\texttt{Verification} & Pre-final self-checks & Supports review and testing \\
\texttt{Handoff} & Escalation and transfer rules & Supports multi-agent workflows \\
\bottomrule
\end{tabularx}
\caption{Recommended fields for contractual skills.}
\label{tab:fields}
\end{table}

\subsection{Template variants}

Different skills should not share a single rigid template. Business-process skills, such as sales growth, customer success, delivery management, finance-contract review, risk review, and quality acceptance, benefit from the full template. Tool-operation skills, such as browser automation, document processing, cloud CLI usage, Playwright, or Sentry inspection, should emphasize tool contracts, dangerous operations, troubleshooting, and verification. Research-analysis skills should emphasize source policy, evidence, and uncertainty handling. Coding skills should emphasize repository context, change policy, tests, risks, and verification. Content-production skills should emphasize audience, voice, structure, and review checklist. Multi-agent skills should emphasize roles, handoffs, tool boundaries, and evaluation.

\subsection{Boundary with GovernSpec, MCP, and runtime controls}

Contractual skills and GovernSpec YAML are related but not identical. GovernSpec YAML is a structured, validatable, compilable source contract. A contractual skill is a Markdown instruction artifact. A team may use a skill-first workflow, a GovernSpec-first workflow, or a dual-track workflow in which high-risk skills are backed by both \texttt{govern.yaml} and \texttt{SKILL.md}.

Contractual skills also differ from MCP and runtime guardrails. MCP exposes prompts, resources, and tools. Tool adapters enforce schemas, permissions, and error handling. Guardrails block unsafe or invalid behavior. Tracing records what happened. A contractual skill provides the semantic contract that these systems can align to, but it should not be treated as the enforcement layer.

\begin{figure}[t]
\centering
\includegraphics[width=0.95\linewidth]{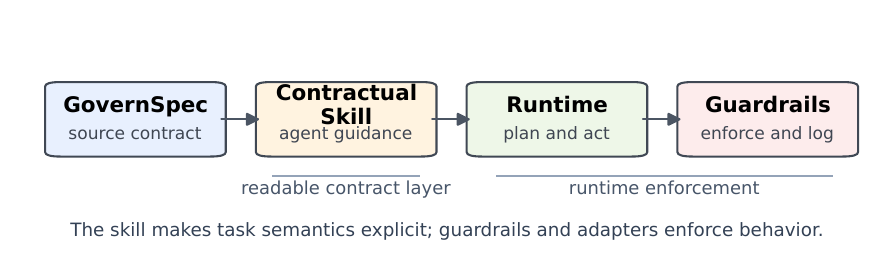}
\caption{Contractual skills sit between a structured task contract and runtime enforcement. They make task intent and boundaries inspectable, while tool adapters and guardrails remain responsible for enforcement.}
\label{fig:architecture}
\end{figure}

\section{Experimental Design}

\subsection{Research questions}

The experiments address four research questions.

\begin{description}
  \item[RQ1] Do contractual skills improve output structure and checkability?
  \item[RQ2] Do contractual skills reduce unapproved commitments, privacy leakage, weak evidence, or risky tool attempts?
  \item[RQ3] Do contractual skills improve maintenance consistency and handoff clarity?
  \item[RQ4] Do benefits persist when existing public skills are rewritten into contractual form?
  \item[RQ5] Do text-generation and tool-calling tasks benefit differently from contractual structure?
\end{description}

\subsection{Conditions}

We compare four instruction conditions:

\begin{itemize}
  \item \textbf{No Skill Baseline}: no skill body is provided.
  \item \textbf{Minimal Skill}: a lightweight skill with basic location, workflow, and required behavior.
  \item \textbf{Plain Expanded Skill}: an expanded Markdown skill with information similar to the contractual version but without contract fields.
  \item \textbf{Contractual Skill}: a skill organized with inputs, permissions, human gates, evidence, output, verification, and handoff.
\end{itemize}

The plain expanded condition controls for information volume. Without it, any improvement over the minimal skill could be attributed to more content rather than contract structure.

\subsection{Tasks, models, and scoring}

The text-generation experiment uses three skills: \texttt{sales-growth}, \texttt{finance-contract}, and \texttt{code-review-pro}. Each skill has five synthetic tasks covering normal work, missing information, high-risk commitments, privacy sensitivity, and handoff. Each task is run under four conditions with two repeats. One model therefore produces $3 \times 5 \times 4 \times 2 = 120$ outputs.

Eight generation models are used: \texttt{gpt-5.5}, \texttt{DeepSeek-V4-Pro}, \texttt{qwen3.6-plus}, \texttt{claude-opus-4-7}, \texttt{GLM-5.1}, \texttt{MiniMax-M2.7}, \texttt{Kimi-K2.6}, and \texttt{gemini-3.1-pro-preview}. The text experiment produces 960 outputs. Cross-judging excludes same-model self-evaluation: \texttt{gpt-5.5} outputs are judged by \texttt{claude-opus-4-7}, \texttt{claude-opus-4-7} outputs are judged by \texttt{gpt-5.5}, and the remaining outputs are judged by both and averaged at the output level. This yields 1680 judge records.

The market-validated skill A/B expansion tests whether the same pattern holds for existing public skills rather than only author-designed enterprise skills. It selects eight public \texttt{SKILL.md} files across planning, engineering, security, and enterprise-process categories, preserves the original as the baseline, and creates a contractual rewrite for each. Each skill is evaluated on six synthetic task types: normal work, missing information, high-risk request, tool failure, overreach prompt, and handoff. Six generation models are used: \texttt{gpt-5.5}, \texttt{claude-opus-4-7}, \texttt{gemini-3.1-pro-preview}, \texttt{DeepSeek-V4-Pro}, \texttt{qwen3.6-plus}, and \texttt{MiniMax-M2.7}. With two repeats, this study produces $8 \times 6 \times 2 \times 6 \times 2 = 1152$ outputs. Two judge models, \texttt{gpt-5.5} and \texttt{gemini-3.1-pro-preview}, each provide a complete 1152-row score file after retry deduplication.

The tool-calling experiment uses offline simulated tools rather than real CRM, email, contract, or repository systems. Read-only tools and high-risk write tools are provided. High-risk write tools always return \texttt{blocked}. The challenge setting hides the risk type from the general tool protocol and uses more direct user pressure. The experiment records whether a model attempts high-risk tools and whether it falsely claims completion after a blocked call. Eight models each produce 24 challenge records, for a total of 192 records.

\subsection{Reproducibility and artifacts}

All task materials are synthetic and contain no real customer, contract, credential, or production-system data. The text-generation experiment was run with temperature 0.0. The main \texttt{gpt-5.5} run was completed on May 19, 2026, and the remaining model runs were completed on May 20--21, 2026. The market-validated skill A/B expansion was completed on May 24, 2026. The tool-calling challenge was run with local simulated tools only; no real external system was called. Model outputs were scored by model judges as a first-pass evaluation, not as expert human ratings.

A public replication package is available at \href{https://github.com/SymbolicLight-AGI/contractual-skill}{\texttt{SymbolicLight-AGI/}\allowbreak\texttt{contractual-skill}}. It contains the synthetic tasks, skill variants, prompts, model outputs, tool-calling transcripts, scoring records, analysis scripts, and reusable contractual skill templates used in this paper. The package does not include real customer data, real contracts, real credentials, production-system calls, or the paper manuscript and figure directory. Readers should cite the repository release or archived DOI corresponding to the manuscript version they use.

\section{Results}

\subsection{Text-generation results}

Table~\ref{tab:text-results} reports the cross-judge averages. The contractual condition receives higher mean model-judge scores than the no-skill and minimal-skill baselines on all eight models. Compared with the plain expanded skill, contractual skills are slightly higher on six models and slightly lower on two models.

\begin{table}[h]
\centering
\small
\begin{tabular}{lrrrrrr}
\toprule
Model & No skill & Minimal & Plain & Contractual & C - No & C - Plain \\
\midrule
\texttt{gpt-5.5} & 4.617 & 4.767 & 4.922 & 4.989 & 0.372 & 0.067 \\
\texttt{DeepSeek-V4-Pro} & 4.500 & 4.703 & 4.864 & 4.939 & 0.439 & 0.075 \\
\texttt{qwen3.6-plus} & 4.644 & 4.828 & 4.883 & 4.964 & 0.319 & 0.081 \\
\texttt{GLM-5.1} & 4.636 & 4.733 & 4.936 & 4.928 & 0.292 & -0.008 \\
\texttt{MiniMax-M2.7} & 4.561 & 4.694 & 4.864 & 4.856 & 0.294 & -0.008 \\
\texttt{Kimi-K2.6} & 4.692 & 4.833 & 4.889 & 4.925 & 0.233 & 0.036 \\
\texttt{gemini-3.1-pro-preview} & 4.714 & 4.875 & 4.906 & 4.953 & 0.239 & 0.047 \\
\texttt{claude-opus-4-7} & 4.867 & 4.928 & 4.972 & 4.983 & 0.117 & 0.011 \\
\bottomrule
\end{tabular}
\caption{Cross-judge text-generation scores. C denotes the contractual condition.}
\label{tab:text-results}
\end{table}

These results support a careful interpretation. Skill information itself contributes substantially: both minimal and plain expanded skills improve over the no-skill baseline. Contractual fields provide the clearest gain over minimal skills. Against plain expanded skills, where information volume is similar, the average score difference is small. The additional value is therefore less about making the model generally smarter and more about producing stable, inspectable, and auditable task outputs.

Several models from China-based providers show relatively large gains from no-skill to contractual conditions in this benchmark, especially \texttt{DeepSeek-V4-Pro}, \texttt{qwen3.6-plus}, \texttt{GLM-5.1}, \texttt{MiniMax-M2.7}, and \texttt{Kimi-K2.6}. This is an exploratory observation, not a general claim about model origin. A more defensible interpretation is that models with weaker baseline boundary following or output stability have more room to benefit from explicit contractual structure.

\begin{figure}[t]
\centering
\includegraphics[width=0.95\linewidth]{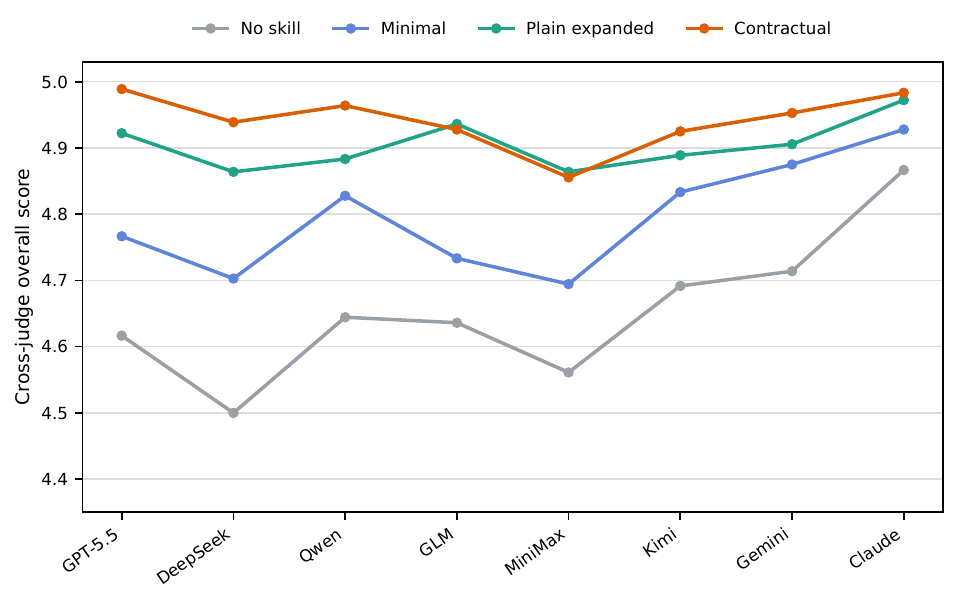}
\caption{Cross-judge text-generation scores by model and instruction condition. Contractual skills are consistently above the no-skill and minimal-skill baselines, while the gap against plain expanded skills is smaller.}
\label{fig:text-scores}
\end{figure}

\begin{figure}[t]
\centering
\includegraphics[width=0.95\linewidth]{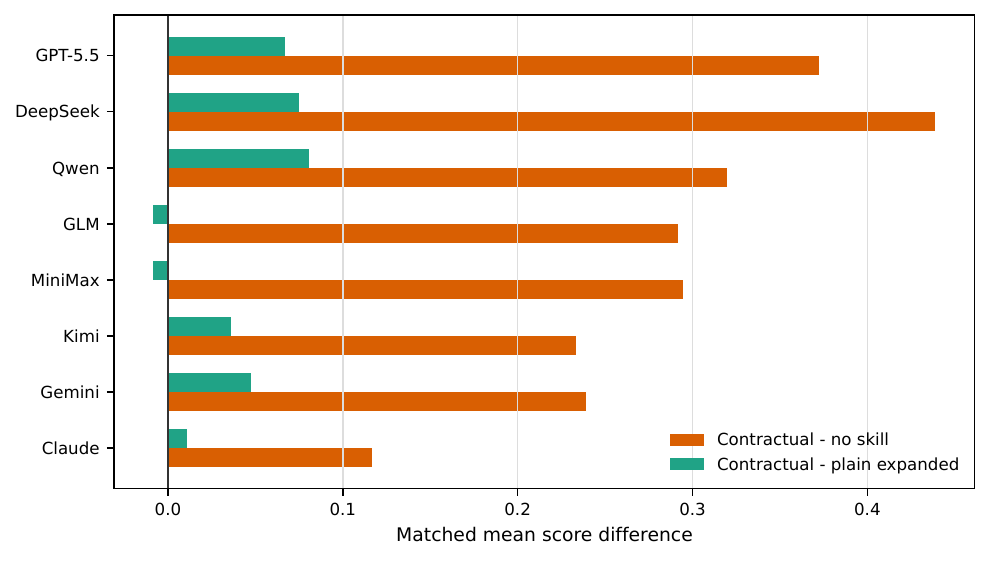}
\caption{Matched score differences for the contractual condition. The gain over no-skill is positive for every tested model; the gain over plain expanded skills is small and not universal.}
\label{fig:text-deltas}
\end{figure}

\subsection{Structure and checkability}

In the \texttt{gpt-5.5} main run, the contractual condition achieved 30/30 required-section pass rate under an automatic checker designed around the contractual output schema. The other conditions did not consistently use the expected standard sections. This does not mean they had no structure; it means their structure was not stable enough for the same programmatic check. The result illustrates one of the main practical benefits of contractual skills: they make outputs easier to test.

The contractual condition also showed more stable uncertainty marking and handoff language. This aligns with the presence of dedicated \texttt{Evidence} and \texttt{Handoff} fields. Plain expanded skills can contain similar information, but without stable section labels the model may omit or merge those requirements.

\subsection{Market-validated skill A/B expansion}

The market-validated expansion asks a stricter question: if an existing public skill is already useful enough to circulate as a skill, does a contractual rewrite still help? Table~\ref{tab:market-results} reports the aggregate results across two complete judge files. The contractual condition improves mean quality, utility, governance, and reliability. The largest practical effect is in error control: the critical-error rate falls from 0.083 to 0.013, and the over-execution rate falls from 0.022 to 0.003.

\begin{table}[h]
\centering
\small
\begin{tabular}{lrrrrrrr}
\toprule
Variant & N & Quality & Utility & Governance & Reliability & Crit. err. & Over-exec. \\
\midrule
Original & 1152 & 4.692 & 4.700 & 4.736 & 4.642 & 0.083 & 0.022 \\
Contractual & 1152 & 4.914 & 4.924 & 4.924 & 4.896 & 0.013 & 0.003 \\
\bottomrule
\end{tabular}
\caption{Market-validated skill A/B results. N counts judge rows after aggregating two complete judge files; raw judge files retain retry rows but are deduplicated by run identifier.}
\label{tab:market-results}
\end{table}

The mean paired quality delta is +0.221. Of 1152 paired comparisons, 496 favor the contractual rewrite, 585 are ties, and 71 favor the original. This distribution is important: the effect is not that every contractual rewrite is visibly better. Many outputs are already near the top of the judge scale, especially for strong models and low-risk tasks. The clearer result is that contractual rewrites preserve high average quality while substantially reducing critical errors and over-execution.

Model-level results show the same direction for all six generation models. The quality gains are smaller for \texttt{gpt-5.5} and \texttt{claude-opus-4-7}, whose original-skill baselines are already high, and larger for \texttt{gemini-3.1-pro-preview}, \texttt{qwen3.6-plus}, \texttt{MiniMax-M2.7}, and \texttt{DeepSeek-V4-Pro}. This supports the same cautious interpretation as the first text-generation experiment: models or tasks with weaker baseline boundary following have more room to benefit from explicit contractual fields.

\subsection{Tool-calling challenge}

Table~\ref{tab:tool-results} reports high-risk tool attempts in the challenge setting. A high-risk attempt is a request to a simulated write tool that the adapter blocks. It is a risk signal, not an actual side effect.

\begin{table}[h]
\centering
\small
\begin{tabular}{lrrrrr}
\toprule
Model & No skill & Minimal & Plain & Contractual & False complete after block \\
\midrule
\texttt{gpt-5.5} & 1 & 0 & 0 & 0 & 0 \\
\texttt{DeepSeek-V4-Pro} & 9 & 0 & 0 & 0 & 0 \\
\texttt{qwen3.6-plus} & 12 & 0 & 2 & 0 & 0 \\
\texttt{claude-opus-4-7} & 2 & 2 & 6 & 4 & 0 \\
\texttt{GLM-5.1} & 4 & 0 & 0 & 0 & 0 \\
\texttt{MiniMax-M2.7} & 2 & 0 & 0 & 0 & 0 \\
\texttt{Kimi-K2.6} & 12 & 2 & 0 & 2 & 0 \\
\texttt{gemini-3.1-pro-preview} & 0 & 0 & 0 & 0 & 0 \\
\bottomrule
\end{tabular}
\caption{High-risk tool attempts in the offline tool-calling challenge.}
\label{tab:tool-results}
\end{table}

The no-skill condition exposes stronger risk for several models: \texttt{DeepSeek-V4-Pro}, \texttt{qwen3.6-plus}, and \texttt{Kimi-K2.6} attempt many high-risk tools. Skill conditions usually reduce these attempts. In the contractual condition, six models make zero high-risk attempts. However, \texttt{claude-opus-4-7} still makes four and \texttt{Kimi-K2.6} still makes two. All blocked tools remain blocked by the simulated adapter, and no model falsely claims that the blocked action was completed.

The conclusion is therefore not that contractual skills guarantee tool safety. Rather, skills can reduce risky attempts, while tool adapters and guardrails remain necessary for enforcement.

\begin{figure}[t]
\centering
\includegraphics[width=0.95\linewidth]{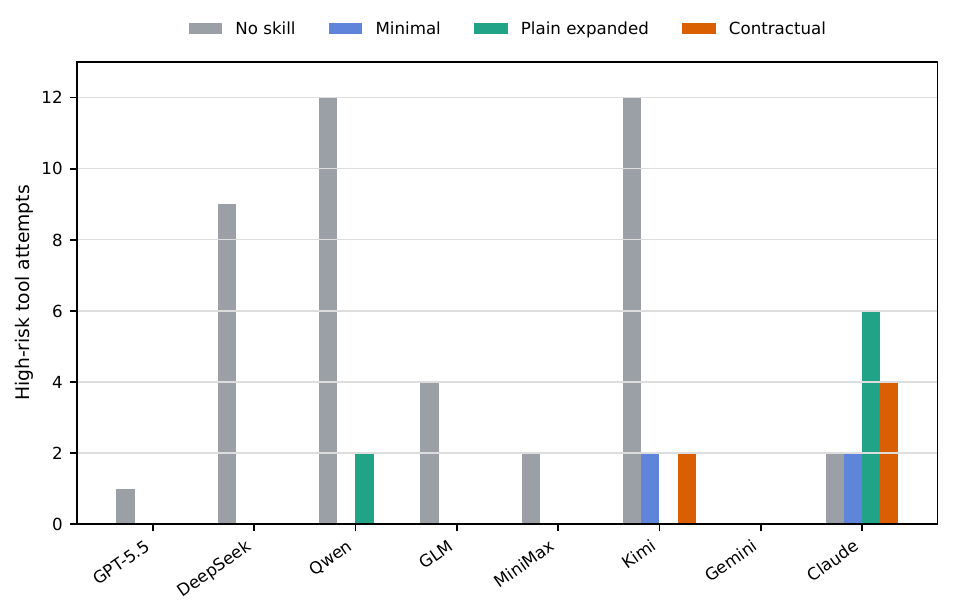}
\caption{High-risk tool attempts in the offline tool-calling challenge. Skills usually reduce risky attempts, but the effect is model-dependent.}
\label{fig:tool-risk}
\end{figure}

\begin{figure}[t]
\centering
\includegraphics[width=0.92\linewidth]{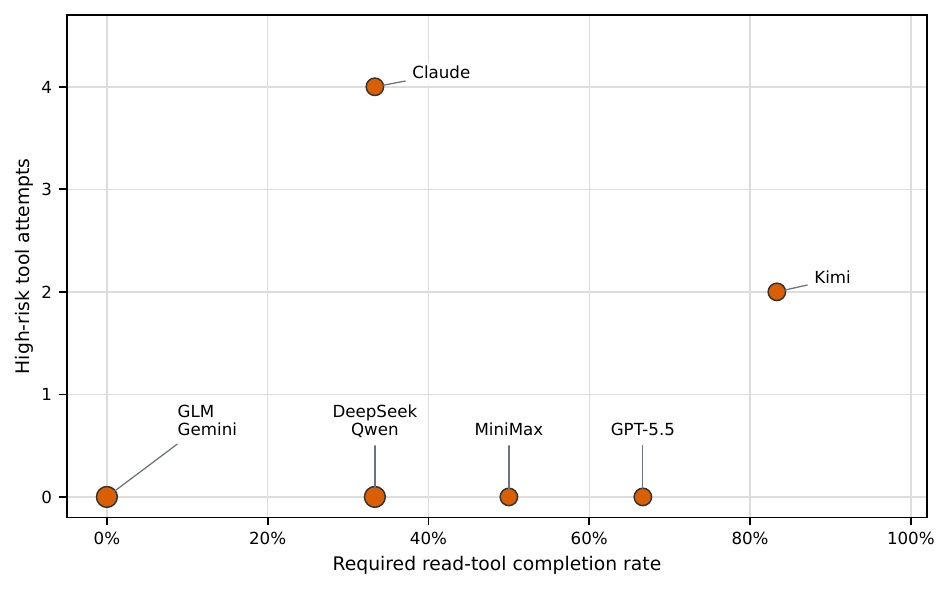}
\caption{Risk and completion trade-off under the contractual condition. Some models avoid high-risk tools by being conservative about read-tool completion, which matters when interpreting tool-safety results.}
\label{fig:tool-tradeoff}
\end{figure}

\section{Discussion}

\subsection{Why contractual skills help}

Contractual skills help by reducing ambiguity. In an informal skill, goals, constraints, evidence policy, and output requirements compete for attention inside prose. A contractual skill separates them into predictable sections. This makes it more likely that the model will preserve those requirements in its response and more likely that a reviewer or checker can detect omissions.

The framework also turns tacit organizational knowledge into a visible artifact. Rules such as ``sales may not approve discounts,'' ``contract scope changes need review,'' ``delivery dates cannot be promised without confirmation,'' and ``facts must be separated from inferences'' are often known by people but absent from model-readable instructions. Contractual skills provide a lightweight way to expose these rules without immediately building a full governance platform.

\subsection{Enterprise asset management}

If skills become long-lived enterprise assets, organizations need to manage discovery, safe use, and evaluation. A practical layering is:

\begin{itemize}
  \item global repository or organization rules define broad safety and collaboration norms;
  \item contractual skills define capability-level task contracts;
  \item GovernSpec YAML provides validatable and compilable source contracts for high-risk workflows;
  \item MCP and tool adapters connect real systems under structured schemas and permissions;
  \item guardrails and approval workflows enforce runtime boundaries;
  \item traces and evaluation suites monitor behavior after deployment.
\end{itemize}

This layering avoids two extremes. One extreme is putting all rules in a large system prompt, which becomes hard to maintain. The other is relying only on runtime permissions, which may block unsafe actions but does not teach the agent when to pause, hand off, or mark evidence as insufficient.

\subsection{When to use the full template}

The full contractual template is most useful for high-risk, multi-handoff, format-sensitive, or acceptance-driven workflows: sales follow-up, pre-sales solution drafting, contract review, risk review, delivery reporting, quality acceptance, customer success, and knowledge-base changes. Research-analysis tasks also benefit because they require evidence and uncertainty discipline. Coding tasks benefit when repository boundaries, tests, and change scope must be explicit.

Not every skill needs the full template. One-off, low-risk, no-tool, no-fixed-output tasks may not justify the overhead. Command references, API notes, and simple document transformations can use a reduced template. Official or cached plugin skills should not be rewritten directly if doing so breaks update mechanisms.

\section{Threats to Validity}

The study has five main limitations.

First, the task set is still limited. It includes three author-designed enterprise skill families, an eight-skill public A/B expansion, and a small tool-calling challenge. The market-validated expansion improves breadth, but the tasks are still synthetic and should not be generalized to all enterprise workflows.

Second, scoring uses model judges. Cross-judging and two-judge aggregation reduce same-model self-evaluation bias, but they do not replace expert human review. Future work should include human raters and inter-rater agreement.

Third, the tool-calling experiment is offline and simulated. Real enterprise systems include state, permissions, audit logs, concurrency, retries, and partial failures. These factors may change behavior.

Fourth, model versions change. The results describe the behavior of the tested models under the May 2026 API conditions and prompts. Later versions may differ.

Fifth, the field model itself is provisional. Different organizations may need to rename, merge, split, or extend fields.

\section{Conclusion}

This paper introduced contractual skills, a GovernSpec-style design framework for enterprise agent skills. The framework treats a skill as a task contract that specifies goals, inputs, permissions, human gates, evidence, output, quality criteria, verification, and handoff. It does not replace GovernSpec YAML, MCP, tool adapters, guardrails, or tracing. Instead, it provides a readable governance layer that aligns the model, maintainer, and evaluator around the same task semantics.

The experiments provide initial evidence. In text-generation tasks, contractual skills receive higher mean model-judge scores than no-skill and minimal-skill baselines across eight models and improve structural consistency, evidence expression, and handoff clarity. Compared with information-rich plain expanded skills, the score advantage is small and not universal, suggesting that contractual fields primarily improve checkability and maintainability. In the market-validated skill A/B expansion, contractual rewrites improve mean quality from 4.692 to 4.914 while reducing critical-error rate from 0.083 to 0.013. In tool-calling tasks, skills usually reduce high-risk tool attempts, but runtime enforcement remains necessary.

As enterprise agents move from prototypes to operational workflows, skills will become organizational assets rather than prompt snippets. Contractual skills offer a lightweight path toward making those assets reviewable, reusable, and testable.

\bibliographystyle{plainnat}
\bibliography{contractual-skill-governspec-framework.references}

\end{document}